\documentclass[aps,prd,twocolumn,showpacs,nofootinbib]{revtex4-1}

\usepackage{graphicx}
\usepackage{amssymb}
\usepackage{color}

\usepackage[colorlinks=true,linkcolor=blue,citecolor=blue, urlcolor=blue]{hyperref}

\begin{document}

\title{Determination of the top-quark $\overline{\rm MS}$ running mass via its perturbative relation to the on-shell mass with the help of the principle of maximum conformality}

\author{Xu-Dong Huang}
\email{hxud@cqu.edu.cn}
\author{Xing-Gang Wu}
\email{wuxg@cqu.edu.cn}
\author{Jun Zeng}
\email{zengj@cqu.edu.cn}
\author{Qing Yu}
\email{yuq@cqu.edu.cn}
\author{Xu-Chang Zheng}
\email{zhengxc@cqu.edu.cn}
\author{Shuai Xu}
\email{shuaixu@cqu.edu.cn}

\affiliation{Department of Physics, Chongqing University, Chongqing 401331, People's Republic of China}

\date{\today}

\begin{abstract}

In the paper, we study the properties of the top-quark $\overline{\rm MS}$ running mass computed from its on-shell mass by using both the four-loop $\overline{\rm MS}$-on-shell relation and the principle of maximum conformality (PMC) scale-setting approach. The PMC adopts the renormalization group equation to set the correct magnitude of the strong running coupling of the perturbative series, its prediction avoids the conventional renormalization scale ambiguity, and thus a more precise pQCD prediction can be achieved. After applying the PMC to the four-loop $\overline{\rm MS}$-on-shell relation and taking the top-quark on-shell mass $M_t=172.9\pm0.4$ GeV as an input, we obtain the renormalization scale-invariant $\overline{\rm MS}$ running mass at the scale $m_t$, e.g., $m_t(m_t)\simeq 162.6\pm 0.4$ GeV, in which the error is the squared average of those from $\Delta \alpha_s(M_Z)$, $\Delta M_t$, and the approximate error from the uncalculated five-loop terms predicted by using the Pad\'{e} approximation approach.

\end{abstract}


\maketitle

\section{Introduction}

In quantum chromodynamics (QCD), the quark masses are elementary input parameters of the QCD Lagrangian. There are three light quarks (up, down, and strange) and three heavy ones (charm, bottom, and top). Comparing with other quarks, the top quark is special. It decays before hadronization, which can be almost considered as a free quark. Therefore, the top-quark on-shell (OS) mass, or equivalently the pole mass, can be determined experimentally. The direct measurements are based on analysis techniques which use top-pair events provided by Monte Carlo (MC) simulation for different assumed values of the top-quark mass. Applying those techniques to data yields a mass quantity corresponding to the top-quark mass scheme implemented in the MC; thus, it is usually referred to as the ``MC mass." Since the top-quark MC mass is within $\sim 1$ GeV of its OS mass~\cite{Buckley:2011ms}, one can treat the MC mass as the OS one~\cite{Khachatryan:2016mqs, Fleming:2007qr, Skands:2007zg, Kawabataa:2014osa, Kieseler:2015jzh}. Detailed discussions on the top-quark OS mass can be found in Refs.\cite{Nason:2016tiy, Beneke:2016cbu, Wang:2017kyd}. As shown by the Particle Data Group~\cite{Tanabashi:2018oca}, an average of various measurements at the Tevatron and the LHC gives the OS mass $M_t=172.9\pm0.4$ GeV.

Practically, one usually adopts the modified minimal subtraction scheme (the $\overline{\rm MS}$ scheme) to do the pQCD calculation, and the $\overline{\rm MS}$ running quark mass is introduced. As for the top-quark $\overline{\rm MS}$ running mass, it can be related to the OS mass perturbatively which has been computed up to four-loop level~\cite{Tarrach:1980up, Gray:1990yh, Chetyrkin:1999qi, Melnikov:2000qh, Marquard:2007uj, Marquard:2015qpa, Marquard:2016dcn, Kataev:2018mob, Kataev:2018sjv}. Using this relation and the measured OS mass, we are facing the chance of determining a precise value for the top-quark $\overline{\rm MS}$ running mass. In using the relation, an important thing is to determine the exact value of the strong coupling constant ($\alpha_s$). The scale running behavior of $\alpha_s$ is controlled by the renormalization group equation (RGE) or the $\beta$ function~\cite{Politzer:1973fx, Politzer:1974fr, Gross:1973id, Gross:1973ju}, which is now known up to five-loop level~\cite{Baikov:2016tgj}. Using the Particle Data Group reference point $\alpha_s(M_Z)=0.1181\pm0.0011$~\cite{Tanabashi:2018oca}, we can fix its value at any scale. And thus the remaining task for achieving the precise value of the perturbative series of the $\overline{\rm MS}$ running mass over the OS mass is to determine the correct momentum flow and hence the correct $\alpha_s$ value of the perturbative series.

Conventionally, people uses the guessed renormalization scale as the momentum flow of the process and varies it within an arbitrary range to estimate its uncertainty for the pQCD predictions. This naive treatment leads to the mismatching of the strong coupling constant with its coefficients, well breaking the renormalization group invariance~\cite{Brodsky:2012ms, Wu:2013ei, Wu:2014iba} and leading to renormalization scale and scheme ambiguities. And the effectiveness of this treatment depends heavily on the perturbative convergence of the pQCD series. Sometimes, the scale is chosen so as to eliminate the large logarithmic terms or to minimize the contributions from high-order terms. And sometimes, the scale is so chosen to directly achieve the prediction in agreement with the data. This kind of guessing work depresses the predicative power of the pQCD theory, and sometimes is misleading, since there may be new physics beyond the standard model.

To eliminate the artificially introduced renormalization scale and scheme ambiguities, the principle of maximum conformality (PMC) scale-setting approach has been suggested~\cite{Brodsky:2011ta, Brodsky:2011ig, Brodsky:2012rj, Mojaza:2012mf, Brodsky:2013vpa}. The purpose of the PMC is to determine the effective $\alpha_s$ of a pQCD series by using the known $\beta$-terms of the pQCD series. The argument of the effective $\alpha_s$ is called the PMC scale, which corresponds to the effective momentum flow of the process. It has been found that the magnitude of the determined effective $\alpha_s$ is independent of any choice of renormalization scale; thus, the conventional renormalization scale ambiguity is eliminated by applying the PMC. The PMC shifts all nonconformal $\beta$-terms into the strong coupling constant at all orders, and it reduces to the Gell-Mann and Low scale-setting approach~\cite{GellMann:1954fq} in the QED Abelian limit~\cite{Brodsky:1997jk}. Furthermore, after adopting the PMC to fix the $\alpha_s$ running behavior, the remaining perturbative coefficients of the resultant series match the series of conformal theory, leading to a renormalization scheme independent prediction. Using the PMC single-scale approach~\cite{Shen:2017pdu}, it has recently been demonstrated that the PMC prediction is scheme independent up to any fixed order~\cite{Wu:2018cmb}. The residual scale dependence due to the uncalculated higher-order terms is highly suppressed by the combined $\alpha_s$ suppression and exponential suppression~\cite{Wu:2019mky}. Because of the elimination of the divergent renormalon terms like $n!\beta_0^n\alpha_s^n$~\cite{Beneke:1994qe, Neubert:1994vb, Beneke:1998ui}, the convergence of the pQCD series is naturally improved, which leads to a more accurate prediction. Moreover, the renormalization scale-and-scheme independent series is also helpful for estimating the contribution of the unknown higher orders, some examples can be found in Refs.\cite{Du:2018dma, Yu:2018hgw, Yu:2019mce}.

\section{Calculation technology \label{II}}

The renormalized mass under the $\overline{\rm MS}$ scheme or the OS scheme can be related to the bare mass ($m_0$) by
\begin{eqnarray}
  m_0 = Z_m^{R} m^{R},
\end{eqnarray}
where $R=\overline{\rm MS}$ or OS. Under the $\overline{\rm MS}$ scheme, one can derive the expression of $Z_m^{\overline{\rm MS}}$ by requiring the renormalized propagator to be finite, which has been calculated up to five-loop level~\cite{Chetyrkin:1997dh, Vermaseren:1997fq, Baikov:2014qja, Chetyrkin:2004mf}. Under the OS scheme, the expression of $Z_m^{\rm OS}$ can be obtained by requiring the quark two-point correlation function to vanish at the position of the OS mass, whose one-, two-, and three-loop QCD corrections have been given in Refs.\cite{Tarrach:1980up, Gray:1990yh, Chetyrkin:1999ys, Chetyrkin:1999qi, Melnikov:2000qh, Marquard:2007uj}, and the electroweak effects have also been considered in Refs.\cite{Hempfling:1994ar, Jegerlehner:2002em, Jegerlehner:2003py, Jegerlehner:2003sp, Faisst:2004gn, Martin:2005ch, Eiras:2005yt, Jegerlehner:2012kn, Kniehl:2015nwa, Martin:2016xsp}. Generally, the relation between the $\overline{\rm MS}$ quark mass and OS quark mass can be written as
\begin{eqnarray}
z_m(\mu_r)=\frac{m(\mu_r)}{M}=\frac{Z_m^{\rm OS}}{Z_m^{\overline{\rm MS}}}=\sum_{n\geq0} z^{(n)}_m(\mu_r) a^n_s(\mu_r),\label{zmmur}
\end{eqnarray}
where $a_s(\mu_r)=\alpha_s(\mu_r)/4\pi$, $m(\mu_r)$ is the $\overline{\rm MS}$ running mass with $\mu_r$ being the renormalization scale, and $M$ is the OS quark mass. The perturbative coefficients $z^{(n)}_m$ have been known up to four-loop level~\cite{Marquard:2015qpa, Marquard:2016dcn}, and the $\overline{\rm MS}$ running mass at the scale $M$ takes the following perturbative form:
\begin{eqnarray}
m(M)&=& M\big\{1+z^{(1)}_m(M)a_s(M) +z^{(2)}_m(M)a^2_s(M) +     \nonumber\\
        &  & z^{(3)}_m(M)a^3_s(M)+z^{(4)}_m(M)a^4_s(M) +\cdots \big\}, \label{zmM}
\end{eqnarray}
where the coefficients $z^{(i)}_m(M)$ $(i=1,\cdots,4)$ can be read from Ref.\cite{Marquard:2016dcn}. Using the displacement relation which relates the $\alpha_s$ value at the scale $\mu_1$ with its value at any other scale $\mu_2$,
\begin{eqnarray}
\label{scaledis}
a_s(\mu_1) = a_s(\mu_2) + \sum_{n=1}^\infty \frac{1}{n!} { \frac{\partial^n a_s(\mu_r)}{(\partial \ln \mu_r^2)^n}|_{\mu_r=\mu_2} (-\delta)^n} \ ,
\end{eqnarray}
where $\delta=\ln {\mu_2^2}/{\mu_1^2}$, and one can obtain the relation at any renormalization scale $\mu_r$, i.e.,
\begin{widetext}
\begin{eqnarray}
m(M)&=&M\Big\{1+ z^{(1)}_m(M)a_s(\mu_r)+\Big[ z^{(2)}_m(M)+\beta _0 z^{(1)}_m(M) \ln\frac{\mu^2_r}{M^2}\Big]a^2_s(\mu_r)+\Big[ z^{(3)}_m(M) +\Big( \beta _1 z^{(1)}_m(M)+2 \beta _0 z^{(2)}_m(M)\Big) \nonumber  \\
&& \ln\frac{\mu^2_r}{M^2}+ \beta _0^2 z^{(1)}_m(M)\ln ^2\frac{\mu ^2_r}{M^2}\Big]a^3_s(\mu_r)+\Big[ z^{(4)}_m(M)+\Big( \beta _2 z^{(1)}_m(M)+2 \beta _1 z^{(2)}_m(M)+3 \beta _0 z^{(3)}_m(M)\Big) \ln\frac{\mu ^2_r}{M^2} \nonumber \\
&&+\Big(\frac{5}{2} \beta _1 \beta _0 z^{(1)}_m(M)+3 \beta _0^2 z^{(2)}_m(M)\Big)\ln ^2 \frac{\mu ^2_r}{M^2}+ \beta _0^3 z^{(1)}_m(M) \ln ^3 \frac{\mu ^2_r}{M^2} \Big]a^4_s(\mu_r)+ \cdots \Big\}. \label{mu}
\end{eqnarray}
\end{widetext}
For the case of top-quark masses, schematically, we can rewrite the perturbative coefficients of the above equation as the $\{n_f\}$-power series,
\begin{eqnarray}
m_t(M_t)&=&M_t\Big\{1+c_{1,0} a_s(\mu_r)+(c_{2,0}+c_{2,1} n_f)a^2_s(\mu_r)  \nonumber\\
& &+(c_{3,0}+c_{3,1} n_f + c_{3,2} n_f^2) a^3_s(\mu_r)  +(c_{4,0}  \nonumber\\
& &+ c_{4,1} n_f + c_{4,2} n_f^2 + c_{4,3} n_f^3 ) a^4_s(\mu_r)+ \cdots \Big\}, \label{cij}
\end{eqnarray}
where $m_t$ is the top-quark $\overline{\rm MS}$ mass and $M_t$ is the top-quark OS mass.

To apply the PMC to fix the $\alpha_s$ value with the help of RGE, we first transform the $n_f$ series as the $\{\beta_i\}$ series by using the degeneracy relations which are the general properties of a non-Abelian gauge theory~\cite{Bi:2015wea},
\begin{eqnarray}
m_t(M_t) &=&M_t\Big\{1+ r_{1,0}a_s(\mu_r) + (r_{2,0}+\beta_{0}r_{2,1})a_{s}^{2}(\mu_r)\nonumber\\
&&+(r_{3,0}+\beta_{1}r_{2,1}+ 2\beta_{0}r_{3,1}+ \beta_{0}^{2}r_{3,2})a_{s}^{3}(\mu_r)\nonumber\\
&& +(r_{4,0}+\beta_{2}r_{2,1}+ 2\beta_{1}r_{3,1} + \frac{5}{2}\beta_{1}\beta_{0}r_{3,2} \nonumber\\
&& +3\beta_{0}r_{4,1}+3\beta_{0}^{2}r_{4,2}+\beta_{0}^{3}r_{4,3}) a_{s}^{4}(\mu_r)+\cdots  \Big\}. ~\label{rij}
\end{eqnarray}
The coefficients $r_{i,j}$ can be obtained from the known coefficients $c_{i,j}$ $(i>j\geq0)$ by applying the basic PMC formulas listed in Refs.\cite{Mojaza:2012mf, Brodsky:2013vpa}. The conformal coefficients $r_{i,0}$ are independent of $\mu_r$, and the nonconformal coefficients $r_{i,j}$ $(j\neq0)$ are functions of $\mu_r$, i.e.,
\begin{eqnarray}
r_{i,j}=\sum^j_{k=0}C^k_j{\hat r}_{i-k,j-k}{\rm ln}^k(\mu_r^2/M_t^2),~\label{rijrelation}
\end{eqnarray}
where the reduced coefficients ${\hat r}_{i,j}=r_{i,j}|_{\mu_r=M_t}$, the combination coefficients $C^k_j=j!/[k!(j-k)!]$, and $i,j,k$ are the polynomial coefficients. For convenience, we put the reduced coefficients ${\hat r}_{i,j}$ in the Appendix.

Applying the standard PMC single-scale approach~\cite{Shen:2017pdu}, the effective coupling $\alpha_s(Q_*)$ can be obtained by using all the nonconformal terms and the perturbative series (\ref{rij}) changes to the following conformal series:
\begin{eqnarray}
&m_t&(M_t)|_{\rm PMC}\nonumber \\
 &=&M_t\Big\{1+ {\hat r}_{1,0}a_s(Q_*) + {\hat r}_{2,0}a_s^2(Q_*)   \nonumber \\
 &  &\quad\quad\quad +{\hat r}_{3,0}a_s^3(Q_*)+ {\hat r}_{4,0}a_s^4(Q_*)+ \cdots \Big\},~\label{conformal}
\end{eqnarray}
where $Q_*$ is the PMC scale, which corresponds to the effective momentum flow of the process and is determined by requiring all the nonconformal terms vanish. The PMC scale $Q^*$, or $\ln{Q^2_*}/{M^2_t}$, can be expanded as a perturbative series, and up to next-to-next-to-leading log (NNLL) accuracy, we have
\begin{eqnarray}
\ln\frac{Q^2_*}{M^2_t}=T_0+T_1 a_s(M_t)+T_2 a^2_s(M_t)+ {\cal O}(a^3_s),
\label{qstar}
\end{eqnarray}
where the coefficients are
\begin{eqnarray}
T_0=&&-\frac{{\hat r}_{2,1}}{{\hat r}_{1,0}}, \\
T_1=&&\frac{ \beta _0 ({\hat r}_{2,1}^2-{\hat r}_{1,0} {\hat r}_{3,2})}{{\hat r}_{1,0}^2}+\frac{2 ({\hat r}_{2,0} {\hat r}_{2,1}-{\hat r}_{1,0} {\hat r}_{3,1})}{{\hat r}_{1,0}^2}
\end{eqnarray}
and
\begin{eqnarray}
T_2=&&\frac{3 \beta _1 ({\hat r}_{2,1}^2-{\hat r}_{1,0} {\hat r}_{3,2})}{2 {\hat r}_{1,0}^2}\nonumber\\
&&+\frac{4({\hat r}_{1,0} {\hat r}_{2,0} {\hat r}_{3,1}-{\hat r}_{2,0}^2 {\hat r}_{2,1})+3({\hat r}_{1,0} {\hat r}_{2,1} {\hat r}_{3,0}-{\hat r}_{1,0}^2 {\hat r}_{4,1})}{ {\hat r}_{1,0}^3} \nonumber \\
&&+\frac{ \beta _0  (4 {\hat r}_{2,1} {\hat r}_{3,1} {\hat r}_{1,0}-3 {\hat r}_{4,2} {\hat r}_{1,0}^2+2 {\hat r}_{2,0} {\hat r}_{3,2} {\hat r}_{1,0}-3 {\hat r}_{2,0} {\hat r}_{2,1}^2)}{ {\hat r}_{1,0}^3}\nonumber\\
&&+\frac{ \beta _0^2 (2 {\hat r}_{1,0} {\hat r}_{3,2} {\hat r}_{2,1}- {\hat r}_{2,1}^3- {\hat r}_{1,0}^2 {\hat r}_{4,3})}{ {\hat r}_{1,0}^3}.
\end{eqnarray}
Using the present known four-loop relations, we can fix the PMC scale up to NNLL accuracy. It can be found that $Q_*$ is independent of the choice of the renormalization scale $\mu_r$ at any fixed order, and the conventional renormalization scale ambiguity is eliminated. This indicates that one can finish the fixed-order perturbative calculation by choosing any renormalization scale as a starting point, and the PMC scale $Q_*$ and hence the PMC prediction shall be independent of such choice.

\section{Numerical results \label{III}}

To do the numerical calculation, we adopt~\cite{Tanabashi:2018oca} $\alpha_s(M_Z)=0.1181\pm0.0011$ and $M_t=172.9\pm0.4$ GeV.

\subsection{Properties of the top-quark $\overline{\rm MS}$ running mass}

\begin{figure}[htb]
\includegraphics[width=0.450\textwidth]{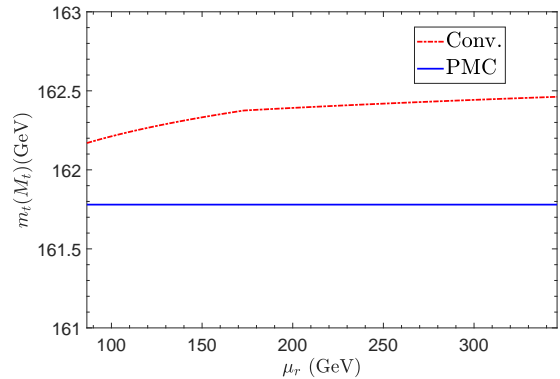}
\caption{The top-quark $\overline{\rm MS}$ running mass, $m_t(M_t)$, up to four-loop QCD corrections under the conventional (Conv.) and PMC scale-setting approaches. The renormalization scale $\mu_r\in[\frac{1}{2}M_t, 2M_t]$. }  \label{mtMt}
\end{figure}

By setting all input parameters to be their central values into Eqs.(\ref{mu}) and (\ref{conformal}), we present the top-quark $\overline{\rm MS}$ running mass at the scale $M_t$ under conventional and PMC scale-setting approaches in Fig.\ref{mtMt}. It shows that the conventional renormalization scale dependence becomes small when we have known more loop terms. Numerically, we obtain $m_t(M_t)|_{\rm Conv.} =[162.170,162.462]$ GeV for $\mu_r \in [\frac{1}{2}M_t, 2M_t]$ GeV, and $m_t(M_t)|_{\rm Conv.} =[162.052,162.522]$ GeV for $\mu_r \in [\frac{1}{3}M_t, 3M_t]$; e.g., the net scale errors are only $\sim0.2\%$ and $\sim0.3\%$, respectively. We should point out that such small net scale dependence for the four-loop prediction is due to the well convergent behavior of the perturbative series, e.g., the relative magnitudes of LO: NLO: N$^2$LO: N$^3$LO: N$^4$LO=1: $4.6\%$: $1\%$: $0.3\%$: $0.1\%$ for the case of $\mu_r=M_t$, and also due to the cancellation of the scale dependence among different orders. The scale errors for each loop term are unchanged and large. For example, the $m_t(M_t)$ has the following perturbative series up to four-loop level:
\begin{eqnarray}
m_t(M_t)|_{\rm Conv.} &=&172.9 - 7.903^{-0.834}_{+0.624} - 1.854^{+0.391}_{-0.276} \nonumber\\
 &  & - 0.560^{+0.175}_{-0.178} - 0.208^{+0.063}_{-0.083}  \; ({\rm GeV})   \nonumber \\
&=&162.375^{-0.205}_{+0.087}  \; ({\rm GeV}),  \label{mtMtconv}
\end{eqnarray}
where the central values are for $\mu_r=M_t$, and the upper and lower errors are for $\mu_r=M_t/2$ and $\mu_r=2M_t$, respectively. It shows that the absolute scale errors are $18\%$, $36\%$, $63\%$, and $70\%$ for the NLO-terms, N$^2$LO-terms, N$^3$LO-terms, and N$^4$LO-terms, respectively.

On the other hand, Fig.\ref{mtMt} shows that after applying the PMC, the relative magnitudes of LO: NLO: N$^2$LO: N$^3$LO: N$^4$LO of the pQCD series changes to 1: $7.2\%$: $0.5\%$: $0.3\%$: $<0.1\%$. And there is no renormalization scale dependence for $m_t(M_t)$ at any fixed order,
\begin{eqnarray}
&  & m_t(M_t)|_{\rm PMC} \nonumber\\
&=&172.9 - 12.497 + 0.919 + 0.551 - 0.095  \; ({\rm GeV}) \nonumber \\
&=&161.778 \; ({\rm GeV}),   \label{mtMtPMC}
\end{eqnarray}
which is unchanged for any choice of renormalization scale. The PMC scale or, equivalently, the effective momentum flow of the process is $Q_*=12.30$ GeV, which is fixed up to NNLL accuracy,
\begin{eqnarray}
\ln\frac{Q^2_*}{M^2_t} &&=-4.686- 51.890 a_s(M_t) - 2126.558 a^2_s(M_t)  \nonumber \\
&&=-4.686-0.445-0.156.
\end{eqnarray}
The relative magnitudes of each loop terms are $1: 9\%: 3\%$, which shows a good convergence. As a conservative estimation, if using the last known term as the magnitude of its unknown NNNLL term, the change of momentum flow is small, $\Delta Q_* \simeq \left(^{+1.00}_{-0.92}\right)$ GeV.

\begin{table}[htb]
\begin{tabular}{ccc}
\hline

      \raisebox {0ex}[0pt]{~~~~}
            & ~~$\rm N^4LO$~~    & ~~$\rm N^5LO$~~  \\
\hline
      \raisebox {0ex}[0pt]{~~$\rm EC$~~}
        & ~~$-0.208^{+0.063}_{-0.083}$~~ & ~~$\cdots$~~ \\
\hline
      \raisebox {0ex}[0pt]{~~$\rm PAA$~~}
       & ~~$[1/1]$; $-0.169^{+0.067}_{-0.086}$~~ & ~~$[1/2]$; $-0.087^{+0.024}_{-0.038}$~~ \\
       & ~~$\cdots$~~ & ~~$[2/1]$; $-0.077^{+0.022}_{-0.038}$~~ \\
\hline
\end{tabular}
\caption{The PAA predictions of the magnitudes of the four-loop and five-loop terms (in unit: \rm GeV)  using the conventional series. The central value is for renormalization scale $\mu_r=M_t$, and the errors are for $\mu_r\in[M_t/2, 2M_t]$. }  \label{convpaa}
\end{table}

\begin{table}[htb]
\centering
\begin{tabular}{ccc}
\hline
      \raisebox {0ex}[0pt]{~~~~}
            & ~~$\rm N^4LO$~~    & ~~$\rm N^5LO$~~  \\
\hline
{~~EC~~}   & ~~$-0.095$~~ & ~~$\cdots$~~ \\
\hline
{~~PAA~~}   & ~~$[0/2]$; $-0.086$~~ & ~~$[0/3]$; $-0.020$~~ \\
\hline
\end{tabular}
\caption{The PAA predictions of the magnitudes of the four-loop and five-loop terms (in unit: GeV)  using the PMC conformal series, which is independent of any choice of renormalization scale. }  \label{pmcpaa}
\end{table}

One usually wants to know the magnitude of the ``unknown" high-order pQCD corrections. We adopt the Pad\'{e} approximation approach (PAA)~\cite{Basdevant:1972fe, Samuel:1992qg, Samuel:1995jc}, which provides a practical way for promoting a finite series to an analytic function, to do such a prediction. It has been found that the conventional pQCD series which has a weaker convergence due to renormalon divergence, the diagonal-type PAA series is preferable~\cite{Gardi:1996iq, Cvetic:1997qm}; for the present case, the $[1/1]$-type and the $[1/2]$-type or $[2/1]$-type are the preferable PAA types to predict the magnitudes of the N$^4$LO- and the N$^5$LO-terms, respectively. And for the more convergent PMC conformal series, the preferred PAA type is consistent with that of the Gell-Mann and Low method~\cite{Du:2018dma} and that of the generalized Crewther relation~\cite{Shen:2016dnq}: e.g., for the present case, the $[0/2]$-type and the $[0/3]$-type are the preferable PAA types to predict the magnitudes of the N$^4$LO- and the N$^5$LO-terms, respectively. More explicitly, following the procedures described in detail in Refs.\cite{Du:2018dma, Wu:2019mky}, we give the PAA predictions of the uncalculated higher-order pQCD contributions in Table~\ref{convpaa} and Table~\ref{pmcpaa} for conventional and PMC scale-setting approaches, respectively. In those two tables, ``EC" stands for the exact results from the known perturbative series, and ``PAA" stands for the PAA prediction by using the known perturbative series: e.g., the N$^4$LO PAA prediction is obtained by using the known N$^3$LO series, etc.

The effectiveness of the PAA approach depends heavily on how well we know the perturbative series and the accuracy of the known perturbative series. Generally, because of large scale dependence for each order terms, the PAA predictions based on the conventional series is not reliable. For the present case, the PAA prediction is acceptable due to the fact that (1) the perturbative series has a good convergence, (2) the large cancellation of the scale dependence among different orders, and (3) the scale dependence of the first several dominant terms are small. More explicitly, Table~\ref{convpaa} shows that by using the conventional pQCD series under the choices of $\mu_r\in[M_t/2, 2M_t]$, the PAA predicted N$^4$LO-term is about $70\%$-$88\%$ of the exact N$^4$LO-term, and the PAA predicted N$^5$LO is about $42\%$-$43\%$ of the exact N$^4$LO-term. On the other hand, the PAA predictions with the help of the renormalization scheme and scale-invariant PMC conformal series is much more reliable. Table~\ref{pmcpaa} shows that by using the PMC conformal series, the PAA predicted N$^4$LO-term is about $91\%$ of the exact N$^4$LO-term, and the PAA predicted N$^5$LO is about $21\%$ of the exact N$^4$LO-term (showing better convergence). Thus, the approximate top-quark $\overline{\rm MS}$ mass up to N$^5$LO level becomes
\begin{eqnarray}
m_t(M_t)|_{\rm Conv.} &=& 162.288^{-0.181}_{+0.049} ({\rm GeV}) \;{\rm [1/2]-type} \\
                                     &=& 162.298^{-0.183}_{+0.049} ({\rm GeV}) \;{\rm [2/1]-type} \\
 m_t(M_t)|_{\rm PMC} &=& 161.758 \; ({\rm GeV}).
\end{eqnarray}

As a final remark, the top-quark $\overline{\rm MS}$ running mass at two scales $\mu_1$ and $\mu_2$ can be related via the following equation~\cite{Baikov:2014qja}:
\begin{eqnarray}
m_t(\mu_1)=m_t(\mu_2)\frac{c_t (4a_s(\mu_1) )} {c_t (4a_s(\mu_2))}, \label{mtrge}
\end{eqnarray}
where the function $c_t(x)=x^{\frac{4}{7}}(1 +1.3980x +1.7935x^2 -0.6834x^3 -3.5356x^4)$. Using Eqs.(\ref{mtMtconv}), (\ref{mtMtPMC}), and (\ref{mtrge}), we obtain the top-quark $\overline{\rm MS}$ running mass at the scale $m_t$:
\begin{eqnarray}
m_t(m_t)|_{\rm Conv.}&&=163.182^{+0.081}_{-0.190}\; ({\rm GeV}),\\
m_t(m_t)|_{\rm PMC}&&=162.629 \; ({\rm GeV}).
\end{eqnarray}

\subsection{Theoretical uncertainties}

After eliminating the renormalization scale uncertainty via using the PMC approach, there are still several error sources, such as the $\alpha_s$ fixed-point error $\Delta\alpha_s(M_Z)$, the error of top-quark OS mass $\Delta M_t$, the unknown contributions from six-loop and higher-order terms, etc. The uncertainty of the four-loop coefficient $z^{(4)}_m(M)$ has been discussed in Ref.\cite{Marquard:2016dcn}, whose magnitude $\sim 0.0004$ GeV is negligibly small. For convenience, when discussing one uncertainty, the other input parameters shall be set as their central values.

As for the $\alpha_s$ fixed-point error, by using $\Delta \alpha_s(M_Z) =0.0011$ together with the four-loop $\alpha_s$-running behavior, we obtain $\Lambda_{\rm QCD,n_f=5}=209.5^{+13.2}_{-12.6}$ MeV and $\Lambda_{\rm QCD,n_f=6}=88.3^{+6.2}_{-5.9}$ MeV. Then we obtain the top-quark $\overline{\rm MS}$ running mass at the scale $m_t$
\begin{eqnarray}
m_t(m_t)|_{\rm Conv.}&=& 163.182^{+0.103}_{-0.103} \;({\rm GeV}), \label{mtmt1}\\
m_t(m_t)|_{\rm PMC} &=& 162.629^{+0.118}_{-0.119} \;({\rm GeV}). \label{mtmt2}
\end{eqnarray}
Eqs.(\ref{mtmt1}, \ref{mtmt2}) show that the PMC prediction is more sensitive to the value of $\Delta \alpha_s(M_Z)$. This is reasonable since the purpose of PMC is to achieve an accurate $\alpha_s$ value of the process, and inversely, a slight change of its running behavior derived from RGE may lead to sizable alterations. Numerically, the determined effective momentum flow $Q_* \simeq 12$ GeV is much smaller than the guessed momentum flow ${\cal O}(M_t)$, and the strong coupling constant is more sensitive to the variation of $\Lambda_{\rm QCD}$.

\begin{figure}[htb]
\includegraphics[width=0.50\textwidth]{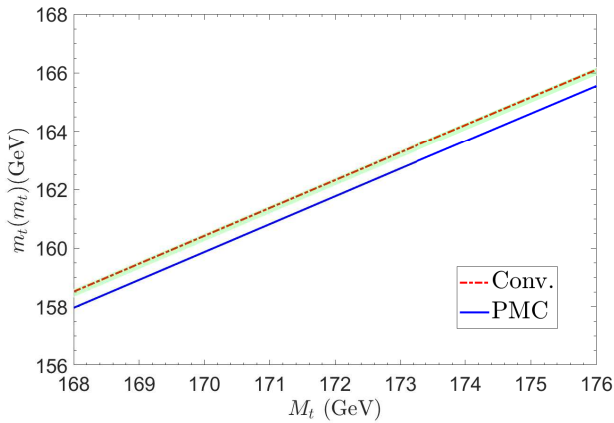}
\caption{The value of the top-quark $\overline{\rm MS}$ running mass $m_t(m_t)$ versus its OS mass $M_t=172.9\pm0.4$ GeV under conventional (Conv.) and PMC scale-setting approaches. The solid line is the PMC prediction, which is independent of the choice of $\mu_r$. The dashed line is the prediction of the conventional scale-setting approach, and the error band shows its errors for $\mu_r\in[M_t/2,2M_t]$, whose lower edge is for $\mu_r=M_t/2$ and upper edge is for $\mu_r=2M_t$.}  \label{mtmtMt}
\end{figure}

As for the error from the choice of the top-quark OS mass $\Delta M_t=\pm0.4$ GeV, we obtain
\begin{eqnarray}
m_t(m_t)|_{\rm Conv.}&&=163.182^{+0.380}_{-0.381} ({\rm GeV}),\\
m_t(m_t)|_{\rm PMC}&&=162.629^{+0.379}_{-0.381} ({\rm GeV}).
\end{eqnarray}
Figure \ref{mtmtMt} shows that the top-quark $\overline{\rm MS}$ running mass $m_t(m_t)$ depends almost linearly on its OS mass, whose error is at the same order of ${\cal O}(\Delta M_t)$.\footnote{As an addendum, if taking the small difference of the OS mass and the Monte-Carlo mass into consideration~\cite{Dehnadi:2018hrh}, $M_t=M_t^{\rm MC}-[0.29\;{\rm GeV}, 0.85\;{\rm GeV}]$, then the above central values of $m_t(m_t)$ shall be altered by about $[0.11, -1.19 ]$ GeV for both conventional and PMC scale-setting approaches.}

In the above subsection, we have predicted the magnitude of the uncalculated N$^5$LO-terms. If treating the absolute value of the PAA predicted N$^5$LO magnitude as a conservative estimation of the error of present N$^4$LO prediction, we shall have an extra error from the unknown perturbative terms, e.g.,
\begin{eqnarray}
\Delta m_t(m_t)|_{\rm Conv.}&=& \pm 0.080 ({\rm GeV}), \;{\rm [1/2]-type} \\
                            &=& \pm 0.072 ({\rm GeV}),  \;{\rm [2/1]-type}\\
\Delta m_t(m_t)|_{\rm PMC}  &=& \pm 0.018 ({\rm GeV}). \;{\rm [0/3]-type}
\end{eqnarray}

\section{summary \label{IV}}

In the present paper, we have presented a more accurate prediction of the top-quark $\overline{\rm MS}$ running mass from the experimentally measured top-quark OS mass by applying the PMC to eliminate the conventional renormalization scale ambiguity. As a combination, we obtain
\begin{eqnarray}
m_t(m_t)|_{\rm Conv.}&=& 163.182^{+0.410}_{-0.445} ({\rm GeV}), {\rm [1/2]-type}\\
                     &=& 163.182^{+0.408}_{-0.444} ({\rm GeV}), {\rm [2/1]-type}\\
m_t(m_t)|_{\rm PMC}  &=& 162.629^{+0.397}_{-0.400} ({\rm GeV}), {\rm [0/3]-type}
\end{eqnarray}
where the errors are squared averages of those from $\Delta \alpha_s(M_Z)$, $\Delta M_t$, and the uncalculated $\rm N^5$LO-terms predicted by using the PAA. Among the errors, the one caused by $\Delta M_t$ is dominant, and we need more accurate data to suppress this uncertainty. The conventional predictions have also the renormalization scale uncertainty by varying $\mu_r \in[\frac{1}{2}M_t, 2M_t]$, even though its magnitude is small due to the cancellation of scale errors among different orders. Up to the present known N$^4$LO level, the predictions under the PMC and conventional scale-setting approaches are consistent with each order. However, it has been found that after applying the PMC, a scale-invariant and more convergent pQCD series, and a more reliable prediction of contribution from unknown higher-order terms can be achieved. Thus, we think the PMC is an important approach for achieving precise pQCD predictions, since its prediction is independent of the choice of renormalization scale. It should be extremely important for lower fixed-order pQCD predictions, when there are not enough terms to suppress the large scale uncertainty of each loop term.

\hspace{0.5cm}

\noindent{\bf Acknowledgements}: This work is partly supported by Graduate Research and Innovation Foundation of Chongqing, China (Grant No. CYB19065), the National Natural Science Foundation of China under Grants No.11625520, No.11947406 and No.11905056, the China Postdoctoral Science Foundation under Grant No. 2019M663432, and by the Chongqing Special Postdoctoral Science Foundation under Grant No. XmT2019055.  \\

\appendix

\section*{Appendix: the PMC reduced perturbative coefficients ${\hat r}_{i,j}$}

In this appendix, we give the required PMC reduced coefficients ${\hat r}_{i,j}$ for the perturbative series of the top-quark $\overline{\rm MS}$ running mass over its OS mass up to four-loop level, i.e.,
\begin{widetext}
\begin{eqnarray}
{\hat r}_{1,0}&=& -4 C_F, \\
{\hat r}_{2,0}&=& C_A C_F \big(6 \zeta_3+5 \pi ^2-\frac{55}{4}-4\pi ^2 \ln 2 \big)+C_F^2 \big(\frac{7}{8}-12 \zeta_3-5 \pi ^2+8\pi ^2 \ln 2 \big)+\big(12-4 \pi ^2\big) C_F T_F, \\
{\hat r}_{2,1}&=& \big(-\frac{71}{8}-\pi^2\big) C_F, \\
{\hat r}_{3,0}&=& C_A^2 C_F \big(51 \pi ^2 \zeta_3+219 \zeta_3-130 \zeta_5-\frac{181 \pi ^2}{6}-\frac{53 \pi ^4}{30}-\frac{19027}{216}+\frac{16 }{3} \pi ^2 \ln 2\big)+C_A C_F^2 \big[384 {\rm Li}_4\big(\frac{1}{2}\big)-76 \pi ^2 \zeta_3 \nonumber \\
&&-112 \zeta_3 +180 \zeta_5 +\frac{518 \pi ^2}{3}+\frac{5731}{12}-\frac{\pi ^4}{15}+16 \ln^4 2-16 \pi ^2 \ln^2 2-\frac{728}{3} \pi ^2 \ln 2\big]+T_F \big[C_A C_F \big(8 \pi ^2 \zeta_3 \nonumber \\
&&+88 \zeta_3 -40 \zeta_5 -\frac{28 \pi ^4}{9}-\frac{4372 \pi ^2}{27}+144+\frac{32}{3}  \pi ^2 \ln^2 2+\frac{1696}{9}  \pi ^2 \ln 2\big)+C_F^2 \big(\frac{56 \pi ^4}{9}-288 \zeta_3-\frac{2608 \pi ^2}{27}\nonumber \\
&&-24-\frac{64}{3} \pi ^2 \ln^2 2+\frac{1216}{9}  \pi ^2 \ln 2\big)\big]+C_F^3 \big[40 \zeta_5 -768 {\rm Li}_4\big(\frac{1}{2}\big)-4 \pi ^2 \zeta_3 -324 \zeta_3-\frac{4 \pi ^4}{3}-\frac{613 \pi ^2}{3}-\frac{2969}{12}\nonumber \\
&&-32\ln^4 2+32 \pi ^2 \ln ^2 2+464 \pi ^2 \ln 2\big]-\frac{608}{45} \pi ^2 C_F T_F^2, \\
{\hat r}_{3,1}&=& C_A C_F \big[32 {\rm Li}_4\big(\frac{1}{2}\big)+\frac{73 \zeta_3 }{2}+\frac{26 \pi ^2}{3}-\frac{19 \pi ^4}{90}-\frac{20335}{432}+\frac{4}{3} \ln^4 2+\frac{8}{3} \pi ^2 \ln^2 2-\frac{44}{3} \pi ^2 \ln 2\big]+ C_F^2  \big[\frac{119 \pi ^4}{90}\nonumber \\
&&-64 {\rm Li}_4\big(\frac{1}{2}\big)-55 \zeta_3 -\frac{95 \pi ^2}{6}-\frac{1927}{48}-\frac{8}{3} \ln^4 2-\frac{16}{3} \pi ^2 \ln^2 2+\frac{88}{3}  \pi ^2 \ln 2\big]+C_F T_F \big(22-24 \zeta_3 -\frac{26 \pi ^2}{3}\big), \\
{\hat r}_{3,2}&=& C_F \big(-14 \zeta_3 -\frac{13 \pi ^2}{3}-\frac{2353}{216}\big), \\
{\hat r}_{4,0}&=& -947.046 C_A^3 C_F-1269.84 C_A^2 C_F^2+T_F \big(3216.18 C_A^2 C_F+568.364 C_A C_F^2-335.759 C_F^3\big)+3671.8 C_A C_F^3\nonumber \\
&&+T_F^2 \big(587.571 C_F^2-2497.16 C_A C_F\big)-\frac{219.883 C_A d_F^{abcd} d_F^{abcd}}{T_F}-1787.65 C_F^4-1050.64 C_F T_F^3+33.28 d_F^{abcd} d_A^{abcd}\nonumber \\
&&-254.891 d_F^{abcd} d_F^{abcd},\\
{\hat r}_{4,1}&=& 932.846 C_A^2 C_F+T_F \big(81.3012 C_F^2-834.037 C_A C_F\big)-473.22 C_A C_F^2-21.945 C_F^3-780.54 C_F T_F^2\nonumber \\
&&+\frac{19.9893 d_F^{abcd} d_F^{abcd}}{T_F}, \\
{\hat r}_{4,2}&=& -112.694 C_A C_F+85.1974 C_F^2-590.868 C_F T_F, \\
{\hat r}_{4,3}&=& -439.436 C_F,
\end{eqnarray}
\end{widetext}
where
\begin{eqnarray}
N_C&&=3,
T_F=\frac{1}{2}, \nonumber \\
C_A&&=N_C,
C_F=\frac{N_C^2-1}{2N_C},\nonumber \\
d_F^{abcd} d_F^{abcd}&&=\frac{(N_C^2-1)(N_C^4-6N_C^2+18)}{96N_C^2}, \nonumber \\
 d_F^{abcd} d_A^{abcd}&&=\frac{N_C(N_C^2-1)(N_C^2+6)}{48}.
\end{eqnarray}

\end{document}